\documentclass[11pt]{article}
\usepackage{bm,bbm,setspace,graphics,amsmath,rotating,amsfonts,amsxtra,amssymb,hyperref,tikz,fullpage}
\usepackage{chngcntr}
\usepackage{apptools}
\usepackage{placeins}
\usepackage{rotating}
\usepackage[utf8]{inputenc}
\usepackage[english]{babel}
\usepackage{natbib}
\usepackage{tcolorbox}
\usepackage{parskip}
\title{Data-NoMAD \\ A Tool for Boosting Confidence in the Integrity of \\ Social Science Survey Data\thanks{This paper describes {\it Data-NoMAD}, which is a copyrighted product of the authors and New York University \copyright~2024. All rights reserved. Data-NoMAD is still in development. If you are interested in trying it out or being a beta tester, please contact the authors.
}}
\author{Sanford C. Gordon \\ Wilf Family Department of Politics \\ New York University \\ sanford.gordon@nyu.edu \and Cyrus Samii \\ Wilf Family Department of Politics \\ New York University \\ cds2083@nyu.edu \and Zhihao Su \\ Center for Data Science \\ New York University \\ zs1512@nyu.edu}

\begin{document}

\maketitle
\begin{spacing}{1}
\begin{abstract}
To safeguard against data fabrication and enhance trust in quantitative social science, we present Data Non-Manipulation Authentication Digest (Data-NoMAD). Data-NoMAD is a tool that allows researchers to certify, and others to verify, that a dataset has not been inappropriately manipulated between the point of data collection and the point at which a replication archive is made publicly available. Data-NoMAD creates and stores a \textit{column hash digest} of a raw dataset upon initial download from a survey platform (the current version works with Qualtrics and SurveyCTO), but before it is subject to appropriate manipulations such as anonymity-preserving redactions. Data-NoMAD can later be used to verify the integrity of a publicly archived dataset by identifying columns that have been deleted, added, or altered. Data-NoMAD complements existing efforts at ensuring research integrity and integrates seamlessly with extant replication practices. 

\end{abstract}
\end{spacing}

\section{Introduction}
\thispagestyle{empty}
A truism concerning empirical research communities is that they rely heavily on trust: a mutually held and accurate belief among members that the underlying data, whether qualitative or quantitative, is authentic. However, a spate of recent revelations of apparent outright fraud in social science research has compelled scholars to reevaluate whether existing approaches to preventing deceptive practices are adequate to ensure that data have not been deliberately altered or even fabricated by a researcher hoping to obtain a favored or flashy result. 

Existing institutions such as peer review, pre-registering studies and hypotheses, pre-analysis plans, the dissemination of replication materials, and results-blind review help guard against inappropriate behavior at the \textit{analysis} stage that threatens, in the aggregate, to undermine the credibility of empirical research \citep{christensen2019transparent, miguel2014promoting, nosek2015promoting, noseklakens2014, brodeur2024preregistration}. 
However, these institutions are less well-suited to dealing with the problem that data are often produced and prepared in ways that lack transparency and are thus susceptible to manipulation and abuse. \textit{Ex post} auditing -- the practice of engaging in forensic-style interrogation of published data -- is the mainline defense against outright data fabrication \citep{broockman2015irregularities, nelson2018psychology}. 
However, the incentives to engage in such audits are weak \citep{clemens2017meaning}.
Auditing does not scale well given the sheer volume of quantitative research, and it can result in both false negatives and false positives. As a consequence, much fraud may go undetected. Additional approaches may be essential for increasing the trustworthiness of, and confidence in, empirical social science research. 

We argue in this note that a useful and effective tool for safeguarding research integrity is to apply \textit{ex ante} measures that facilitate ex post authentication of data. Specifically, we introduce a web-based application called Data Non-Manipulation Authentication Digest, or Data-NoMAD for short. Data-NoMAD employs authentication technology (specifically, the SHA-256 hash function) that, when combined with researcher best practices, will allow researchers to certify the veracity of their data and permit third-party researchers to verify the integrity of archived datasets. Data-NoMAD operates in two modes:

\begin{enumerate}
    \item In \textit{Digest} mode, the application creates cryptographic hash values for all columns in a dataset upon initial download from a survey platform, and stores these values in a lookup table along with column names and a unique survey identifier. The researcher is then free to delete columns containing identifying information before saving an otherwise identical version of the collected data. Any subsequent manipulations of the data would be done in the context of a data preparation coding script available in replication materials along with the de-identified version of the original, raw data.
    
    \item In \textit{Verify} mode, a third party user feeds Data-NoMAD the survey identifier and archived dataset. Data-NoMAD generates a report identifying the names of any columns that were deleted, added, or altered. 
\end{enumerate}

With one caveat, which we discuss below and for which we are currently developing workarounds, Data-NoMAD thus identifies ways in which a dataset has been altered or tampered with between the collection and analysis stages. Typically, a ``clean'' version of the data would produce a verification report in which no columns were flagged as altered and only columns pointing to identifying information are flagged as deleted. 

When combined with researcher best practices, which we discuss below, Data-NoMAD thus mitigates a number of potential avenues for fraud, including manipulated responses, column addition and deletion, and row addition and deletion.  Importantly, and with this in mind, Data-NoMAD is intended to \textit{complement}, rather than substitute for, existing frameworks for ensuring the integrity of scholarly research. These include the posting of replication materials as a condition for publication \citep{king1995replication}; pre-registration \citep{christensen2019transparent} and the ``RARE'' (Reporting All Results Efficiently) framework \citep{laitin2021reporting}; and best practices for data management (e.g., \citealp{gentzkow2014code}).  

In what follows, we describe the context of research fraud that creates the need for Data-NoMAD, and why our strongest current defense against fraud -- ex post auditing -- may be inadequate to the task of policing against deceptive practices. Next, we describe the typical workflow of a quantitative social science researcher employing novel survey data, potential dishonest practices, and \textit{legitimate} data manipulations (most importantly, redaction to preserve the anonymity of human subjects) that must be permitted. We then describe Data-NoMAD in greater detail -- what it does and how it works, its computation and storage architecture, remaining vulnerabilities, and best practices for researchers employing the application.

No system for preventing fraud is perfect. With that said, Data-NoMAD advances a number of goals to the benefit of the social science community. Most immediately, by identifying illegitimate data manipulations, Data-NoMAD helps honest scholars \textit{detect} fraudulent scholarship. Second, by significantly raising the cost of engaging in deceptive research practices, the application aims to \textit{deter} deceptive practices in the first place. Finally, Data-NoMAD helps honest researchers \textit{distinguish} themselves from dishonest ones. The net result, we hope, is to facilitate greater confidence in the veracity of social science data.

\section{Background}

\subsection{Context}

The context of this proposal is a spate of recent, highly publicized instances of outright data fabrication in published social science research. Some of the most prominent recent accusations concern:

\begin{itemize}
\item a social psychologist at Tilburg University, who may have fabricated data for as many as 58 published papers
\item a political science graduate student, accused of fabricating data used in a prominent and subsequently retracted article in \textit{Science} 
\item A social psychologist whose data was found to have suspiciously duplicated observations and unnaturally distributed outcome measures
\item A business school professor implicated in data fabrication in at least four independent papers
\end{itemize}

Importantly, these are cases of outright data fabrication, and hence differ in important ways from other kinds of manipulation at the \textit{analysis stage}: p-hacking, the file drawer problem, abuse of researcher degrees of freedom, hypothesizing after results are known (HARKing), etc. For starters, those practices are often more the product of the fact that researchers are human and subject to biases and motivated reasoning like everyone else, and less the product of a deliberate intent to deceive. But also, a number of institutions aimed at mitigating these issues are already in place: peer review is the oldest and most obvious one; more recent innovations include requirements for the dissemination of replication materials, pre-registration, and results-blind review. 

These recent scandals, however, point to a weak link: data produced and prepared in ways that are non-transparent and thus open to potential manipulation and fabrication. While the incidence of outright fraud is unknown, there are several reasons to believe that it may be more widespread than the cases that we know about. For one thing, splashy research findings may pay enormous dividends in terms of wealth, power, and disciplinary prestige. Conversely, as we discuss in greater detail below, the incentives to invest in uncovering fraud are quite weak. Moreover, the cases we know about have generally come to light because of some sloppy behavior on the part of the researcher. A more sophisticated bad actor might take pains to avoid the kinds of telltale signs of dishonest data manipulation -- to wipe down the fingerprints, as it were.

Especially in the United States, highly publicized cases of research fraud take place in a fraught political environment for the research enterprise in general. Recent public opinion polls suggest declining trust of ordinary citizens in higher education, particularly since the onset of the COVID-19 pandemic, with distrust particularly pronounced among Republicans \citep{brenan_gallup_2023}. While this distrust is surely driven to a great extent by significant ideological differences between researchers and the public, it is surely compounded by a belief that research has been corrupted by outright dishonesty \citep{barro_2024}.

\subsection{Inadequacy of Ex Post Auditing}

In the social and natural sciences, the main approach for curtailing abusive practices is \textit{ex-post auditing}: forensic-style replication that digs into irregularities in a researcher's data by looking for the fingerprints of manipulation or fabrication. These forensic replications complement the more traditional kinds of replication analyses that have become prominent in the 21st century, in which scholars interrogate the choices researchers made on the path between data collection to the presentation of results, but implicitly assume no chicanery in the construction of the underlying datasets themselves.\footnote{Important intermediate cases are replications by \cite{herndon_etal_cje_2013} of \cite{reinhart_rogoff_aer_2010}; by \cite{grimmer_etal_jop_2018} of \cite{hajnal_etal_jop_2017}; and by \cite{fowler_hall_jop_2018} of \cite{achen_bartels_2016}. These replications uncovered errors (some of which were disputed by the original authors), but not deliberate manipulations.}

Ex-post auditing is a vital part of the social science enterprise, and we applaud the scholars who have devoted a considerable portion of their time to uncovering dishonest practices. These efforts provide enormous social benefits to the scholarly community and society at large. At the same time, however, the forensic approach has several serious limitations. First, to the extent that these exercises confer few professional benefits and may even (if the target is a powerful, prestigious member of a discipline) come with significant professional risks, efforts aimed at uncovering fraud are likely to be underprovided relative to the social optimum. This problem is compounded by the fact that uncovering abuse is a public good, i.e., one whose benefits are enjoyed by contributors to that good and non-contributors alike; hence, it is susceptible to a free-rider problem. Compounding this issue still further is the sheer volume of research -- auditing all of it for data manipulation would seem an impossible task. 

A second issue with the forensic approach is that when fraud is uncovered, the evidence for that fraud contains within it a recipe for thwarting detection in the future. For example, in their audit of a 2014 paper by LaCour and Green, \citep{broockman2015irregularities} found that the distribution of baseline thermometer scores in the data used in the Lacour and Green paper was more or less identical to the distribution of scores from the 2012 Cooperative Campaign Analysis Project (CCAP) survey and that the distribution of thermometer scores from subsequent waves did not demonstrate the pronounced ``heaping'' at 0, 50, and 100 generally associated with thermometer scores -- almost certainly because those subsequent scores were manufactured by combining scores from the (fictitious) baseline wave with normally-distributed random noise. Any future would-be fraudsters now know that to reduce the odds of detection they should add noise to data in a way that obscures the provenance of the fictitious data while preserving the heaping pattern. Another example would be the recent analysis of a 2012 paper by Shu, Mazar, Gino, Ariely, and Bazerman by \citet{simmons2021evidence}. They find evidence of manipulation in the calcchain.xml file that is part of a Microsoft Excel workbook. Any bad actor familiar with the case now knows that exporting from Excel to a .csv file and posting the latter in a replication archive can obscure evidence of manipulation. 

The preceding issues suggest the possibility of Type II errors in fraud detection -- instances where actual past, current, or future fraud will go undetected. A third issue with auditing research ex-post concerns Type I errors -- honest research erroneously accused of being fraudulent. The damage to a scholar's reputation associated with a wrongful accusation could be crippling and irreversible. Moreover, there is no guarantee that someone motivated for reasons of professional advancement to commit fraud might not also be motivated to engage in fraudulent accusations for related reasons.\footnote{To be clear, we are absolutely \textit{not} saying that this is the case for the whistleblowers mentioned above, rather that it is not inconceivable that it could be the case in the future.} The flip side of this is that suspicion of impure motives on the part of scholarly whistleblowers deters the enterprise of fraud detection in the first place. More generally, we have no \textit{a priori} reason to believe that the extant incentive structure either minimizes or balances Type I and Type II errors in a socially optimal way.

\section{Dissecting the Problem in Greater Detail}

In this section, we begin by discussing the typical workflow for survey-based research and various points of vulnerability. We then proceed to discuss five different ways that a nefarious researcher could manipulate data in a way that generates fraudulent research and that would generally be undetectable by the research community. Lastly, we discuss \textit{legitimate} reasons why one would make changes to data that any research integrity tool must safeguard.

As mentioned above, the issues we discuss do not include problematic practices that existing norms (e.g., pre-registration and replication), when scrupulously followed, can prevent or discourage. These include, but are not limited to, $P$-hacking, specification search, HARKING (hypothesizing after results are known), and strategically stopping data collection at an advantageous moment. 

\subsection{Typical Workflow and Vulnerabilities}

\begin{figure}
\caption{\label{fig:typical_worfklow}The ``Chain of Custody'' of the Typical Survey Dataset}
\centering
    \includegraphics[width=0.3\linewidth]{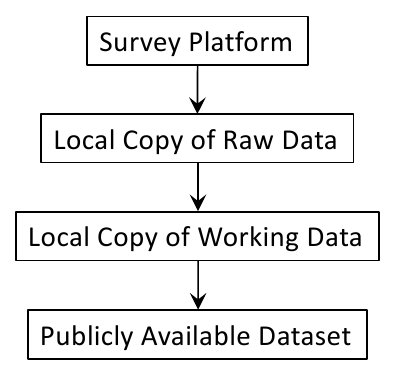}
\end{figure}

The flowchart depicted in Figure \ref{fig:typical_worfklow} displays the typical trajectory of a survey-based dataset assembled for a social science study. First, the research conducts the survey via a platform like Qualtrics, SurveyCTO, SurveyMonkey, or similar. 
Ostensibly, the researcher collects data in line with scientific principles, although we recognize the potential for manipulation at the data collection stage, as we discuss below on adding rows to the data.
Once the data are collected, a dataset is downloaded by the researcher to the researcher's local machine (and perhaps backed up on a cloud service like Dropbox). The researcher may conduct manipulations (legitimate or otherwise) on the local copy to create a ``working'' version of the data, on which statistical analysis is conducted. Finally, and generally in conjunction with the publication of results, some version of the working dataset is made publicly available, for example on a replication archive like Dataverse, the journal's website, or the researcher's website. 

It is not difficult to see that vulnerabilities exist at all points in the ``chain of custody.'' Depending on the platform, there may be options to edit or delete responses before downloading. The researcher is also in a position to make alterations to the local copy of the raw data in various ways that we document below, make changes to the raw data that appear in the working data, or make changes before publicly releasing the data. Critically, changes made to the dataset at any stage propagate to all subsequent steps in the chain of custody. 

\subsection{Potential Deceptive Practices}

We now describe five ways in which data can be manipulated deceptively or fraudulently to contrive an interesting but fictional finding. To summarize, specific entries in a dataset can be edited, and entire rows or columns in a dataset can be removed or added. 

To illustrate how the data can be manipulated deceptively or fraudulently, we will rely on the following fictional political behavior example: A survey research team hypothesizes that exposure to negative advertisements suppresses political participation. The researcher randomly assigns 100 subjects to a treatment condition (a negative political advertisement) or a control condition (an advertisement for a clothing company) and then records a response to the question, ``Do you intend to participate in the upcoming election (Yes/No)?'' 48 respondents are put in the control group and 52 in the treatment group. 32 respondents exposed to the control (67\%) answer yes, while 40 respondents in the treatment group (77\%) answer yes. The difference (10 percentage points) is in the hypothesized direction. However, using conventional hypothesis testing, the researcher would be unable to reject the null hypothesis of no difference between the treatment and control groups ($p=0.26$, two-tailed). 

The researchers wish to manufacture a statistically significant result to enhance the prospects for publication. What can they do?

\paragraph{Response editing.} The most obvious way to proceed would be to simply change the recorded outcomes for individuals in the treatment and/or control groups. In our fictional example, the researchers could switch four of the yes responses to no responses in the control group. Doing this decreases stated vote intention in that arm from 67\% to 58\%, leading to a statistically significant difference of 19 percentage points ($p=0.047$, two-tailed). 

\paragraph{Row deletion.} A second way the researchers might manipulate the data is to delete rows that are inconvenient for their analysis. In our example, inconvenient rows include yes responses in the treatment group and no responses in the control group. The researchers could delete three ``yes'' rows from the control group and four ``no'' rows in the treatment group. This reduces the sample size from 100 to 93, but increases the statistical significance of the reported difference (again 19 percentage points) to a level that permits rejection of the null ($p=0.038$, two-tailed).

\paragraph{Column deletion/replacement.} A third way the researcher can manipulate the data is to simply delete an inconvenient column. In our example, the researchers could outright delete the outcome or treatment columns and replace them with completely fabricated data. 

Another reason that a researcher might delete a column is if it contains an inconvenient confounder. For example, suppose a researcher hypothesized that a measure of authoritarian personality predicted attitudes toward affirmative action, but that the statistically significant relationship disappeared once one controls for partisanship. By deleting the partisanship column the researcher can prevent third parties from testing whether the observed effect of the authoritarianism measure is robust to the inclusion of the partisanship control.

\paragraph{Row addition.} A fourth way to change the data in a way that tilts in the direction of a significant result would be to add fictional responses. Returning to our example, the researchers can generate a statistically significant finding by adding nine new rows to their dataset: four fictional subjects in the control group that answered no, and five fictional subjects in the treatment group that answered yes. This yields a statistically significant difference of 17 percentage points ($p=0.047$, two-tailed).

Another way to add rows is to instruct a confederate to take the survey multiple times, entering values that bolster evidence for the researcher's hypothesis. It could also be the case that the same respondent not affiliated with the study figures a way to take the survey multiple times, or that the respondent isn't a person at all but a bot. Bogus responses on surveys could be inadvertent: for example, investigators might ask their research assistants or students to take the survey multiple times to check for user interface issues, and then accidentally include those responses in the final analysis.

\paragraph{Column addition.} A final deceptive approach would be to add a new fictional variable to the dataset. There are two reasons that a researcher might do this. First, by adding a new variable that is negatively correlated with the treatment of interest and positively correlated with the outcome, the researcher can boost the statistical significance of the effect of the treatment in a regression analysis that contains both old and new variables. This may be particularly tempting if the treatment alone is close to a significance threshold. In our example, we were able to achieve a $p$-value of $0.014$ on the treatment by generating thousands of dichotomous random variables negatively correlated with the treatment until we found one that ``did the trick'' when we included it in a regression.

Second, the researcher might select a subset of observations from the treatment and control groups that generate a significant result, and create a bogus new variable with a plausible-sounding label -- say, blue eye color -- equal to one for those observations and zero otherwise. The researcher could then report the null finding for the average treatment effect but claim heterogeneous effects by eye color. 

\subsection{Legitimate Manipulations}

The list of fraudulent actions above represents a compendium of deceptive practices against which we wish to safeguard. However, as is well known to honest researchers, there are many legitimate reasons analysts will alter their data. Indeed, research ethics may \textit{demand} such changes, for example, to safeguard subject anonymity. Any research integrity tool must permit these practices while discouraging deceptive ones.

\paragraph{Redaction for anonymity.} Completed surveys may include fields that would be unethical or otherwise inappropriate to include in a publicly available replication dataset. These might include respondent contact information (name, address, email address, phone number), IP address, or latitude and longitude.\footnote{Below we discuss an extension of Data-NoMAD that generates a field of unique codes associated with unique IP addresses.}

\paragraph{Data transformations.} The researcher may wish to transform data for analysis purposes: converting text responses to numeric codes; performing mathematical transformations such as logging a variable; or summing numerical responses across multiple fields.

\paragraph{Deletion of extraneous fields.} Survey platforms may report a host of fields of no value to the researcher, such as the distribution channel or user language. 

A more common use case involves situations in which a block of questions that are specific to a particular project ``piggybacks'' on a larger survey. In such circumstances, the researcher may reasonably prefer not to include questions from other parts of the survey in a replication archive. 

\paragraph{Deletion of problematic responses.} The researcher can have legitimate reasons for deleting certain responses.  The raw data may contain both incomplete and completed responses and the researcher might want to conduct analysis only on the completed responses. There might be responses that the researcher suspects were conducted by a bot or a subject who was engaged in button mashing (for example, if the time to completion is a massive outlier). Or perhaps the first 20 responses were test responses by the researchers themselves or their assistants.
Researchers may be committed (perhaps by IRB requirement) to omit responses {\it ex post} if subjects opt out following a debrief or if the researcher learns that certain subjects were ineligible.  
It would be appropriate in such circumstances to exclude those responses from the analysis.

\paragraph{Data joins.} The researcher may wish to join a survey dataset with external data to test for the effects of contextual factors. For example, suppose the researcher is asking questions about anti-immigrant sentiment. The researcher might merge the survey data with auxiliary data on the prevalence of foreign-born residents in the respondents' neighborhoods. 

Another type of legitimate data join is row concatenation: for example, perhaps two separate surveys contain a block of identical questions and the researcher wishes to pool them for analysis.

\section{Data-NoMAD}

\subsection{What it does}

At present, Data-NoMAD operates as a web-based portal that uses authentication technology that, when combined with researcher best practices, allows researchers to certify the credibility of data they have collected and permit third party researchers to verify the integrity of archived datasets.  The authentication technology is the ``Secure Hash Algorithm 256'' (SHA-256) cryptographic security method \citep{fips180-2}.

A hash function is an algorithm that maps data of any arbitrary size to a string of fixed character length (a ``digest''). For example, consider the string
\begin{center}
    The quick brown fox jumps over the lazy dog.
\end{center}
The SHA-256 hash value for this text string is
\begin{center}
    ef537f25c895bfa782526529a9b63d97aa631564d5d789c2b765448c8635fb6c.
\end{center}
Critically, it is computationally infeasible to recover the original data from the hash value. Hash functions are used in innumerable applications from cryptography (including cryptocurrency) to password storage.  

The application, which at present works in conjunction with the Qualtrics and SurveyCTO platforms, works as follows:

\paragraph{Digest Mode (to be used by the researcher).} Digest mode is employed by the researcher after the survey window closes but before the researcher makes any changes to the data for analysis.
\begin{itemize}
    \item The researcher obtains an API key from, e.g., Qualtrics 
    \item The researcher enters the API key and other survey-identifying information into the portal
    \item The portal obtains a temporary copy of an unedited version of the collected data
    \item The portal creates an SHA-256 hash of each column of the data and stores it in a lookup table that contains a unique survey identifier, the name of the column, and the column's 256-bit hash\footnote{Technically, the portal takes two intermediate steps prior to the SHA-256 hash step to more efficiently process the information contained in a column of data: first, it creates a numeric hash array for entries in the column; and second, it converts the array into a contiguous byte sequence. It is this byte sequence whose SHA-256 hash value is stored.} 
    \item Optional (Qualtrics only): the portal creates a hash for each IP address contained in the data
    \item The portal saves a .csv version of the collected data on the user's local machine that (if the IP option is used) contains the hashed IP addresses as a new column. 
\end{itemize}
The researcher would then be free to delete columns containing identifying information before saving an otherwise identical version of the collected data. Any subsequent manipulations of the data would be done in the context of a recoding script available in replication materials along with the de-identified version of the collected data.

It is essential to note that \textit{the raw data from the survey platform is not stored on the Data-NoMAD server}. Thus, using Data-NoMAD poses no more risk of revealing subject identities (thus violating human subjects protections) than the act of downloading the data directly would. 

The IP address hash option bears some discussion. An important potential source of fraud is a single respondent or group of respondents (whether confederates of the researcher or not) taking the survey multiple times. If each respondent on the survey has a unique IP address, this significantly mitigates the concern that the data were improperly collected. The challenge is that IP addresses are potentially identifying information. The hash option allows the analyst to retain a column of alphanumeric codes each corresponding to a unique IP address in the data, while discarding the IP addresses themselves. The 256-bit hash of \textit{this} column of hash values is stored in the Data-NoMAD database along with the hashes of the other columns.\footnote{Technically, the hash takes as its input the IP address and some random text. This is done to prevent a bad actor from cracking the IP hashes by brute force by hashing every possible IP address and comparing them with the values in the column.}

\paragraph{Verify Mode.} Verify mode can be used by other researchers to authenticate a replication dataset; it can also be used by the original investigator to make sure no changes to the data were accidentally introduced during the recoding or analysis stages.
\begin{itemize}
    \item The third party obtains the survey identifier and the deidentified version of the collected data
    \item The third party uploads the data to the portal with the survey identifier
    \item Data-NoMAD produces a report consisting of the following information
    \begin{itemize}
        \item The names of any columns that were deleted
        \item The names of any columns that were added
        \item The names of any columns that were altered (including the IP hash column)
    \end{itemize}
\end{itemize}
A ``clean'' version of the data would have a report in which no columns were altered and only columns pointing to identifying information had been deleted. 

\paragraph{An example}

We next present a deliberately simplistic example of how Data-NoMAD works. An analyst asks a group of individuals for their first names, favorite color, age, and whether they prefer coffee or tea. The Qualtrics ID for the survey is SV\_0q9y0TA1fUsvrkG, and the relevant columns from a Qualtrics .csv file look like this:

%\singlespacing
\begin{center}
\begin{tabular}{|c|c|c|c|}
\hline
name &	color	& age	& drink \\ \hline
What is your  &	What is your &	How old ?&	Do you prefer \\
first name? &favorite color? & are you & coffee or tea? \\ \hline
\{"ImportId": & \{"ImportId": & \{"ImportId": &	\{"ImportId": \\
"QID1\_TEXT"\} & "QID2\_TEXT"\} & "QID3\_TEXT"\} &"QID4"\} \\ 
\hline
Robert	&green	&52	&coffee\\ \hline
Jane	&Topaz	& 27 &	tea \\ \hline
Herman	&Chartreuse&	36	&tea \\ \hline
$\vdots$&$\vdots$&$\vdots$&$\vdots$\\ \hline
Frida	&taupe	&19	&coffee \\ \hline
\end{tabular}
\end{center}
%\doublespacing
The file contains columns of response-specific or survey-specific metadata (e.g., StartDate, EndDate, and IPAddress), and other reserved fields that are frequently left blank (e.g.,  hookType, hookParams). 
Survey platforms such as Qualtrics and SurveyCTO can generate .csv files with column meta-data, obviating the need to work with data formats such as Stata .dta or R .RData files.

The analyst uses Data-NoMAD to download a local .csv file. Data-NoMAD saves the survey ID, column names, and column hashes for all fields in the data (including the column of hashed IP addresses):

%\singlespacing
\begin{center}
    \begin{tabular}{|c|l|l|}
\hline
survey\_id & column\_name & column\_hash \\ \hline
SV\_0q9y0TA1fUsvrkG & StartDate & 1c01f6f7d24c54f81b2eefbc5cbb7b50 \\ 
 & & 2233ca2a2d031abec3f90ea41155128c \\ \hline
 $\vdots$ & $\vdots$ & $\vdots$ \\ \hline
 SV\_0q9y0TA1fUsvrkG & IPaddress & 7894eafa3f9cae1a19048dae4a2982c7 \\ 
 & &f696ea6071304a620a6bc906515c525f \\ \hline
 SV\_0q9y0TA1fUsvrkG & name & 96c6304ae92f64fb8c6eea32665b16d \\
 & & 430539ac3cf169fdc6ce2fc96baf9fa5b \\ \hline
 SV\_0q9y0TA1fUsvrkG & color & 527da150cdfd38bc7c12709da41d1cb \\
 & & 08ec9267632d5298b3ccd027ba14e92fe \\ \hline 
 $\vdots$ & $\vdots$ & $\vdots$ \\ \hline
\end{tabular}
\end{center}
%\doublespacing
The application then saves the .csv file (potentially with an added column of IP address hashes) on the user's local machine (and, to reiterate, \textit{not} on its own server). 

Suppose the researcher (appropriately) deletes the identifying columns \textit{IPaddress} and \textit{name} as well as the extraneous blank ones (a practice we advise against -- see below). But he also changes Robert's favorite color from green to brown. He saves the altered dataset as a new file and, following publication, makes it publicly available. Another researcher wishes to use the data and uploads the public file to Data-NoMAD using ``Verify'' mode. Data-NoMAD would report the following:

\begin{tcolorbox}[width=\linewidth, sharp corners=all, colback=white!85!red]
{\sffamily
%\singlespacing
\noindent Changes detected.
\begin{itemize}
    \item Removed columns: IPaddress, hookParams, name, RecipientID, hookType, PanelID, EmbeddedData, PanelMemberID,
    \item Modified columns: color
\end{itemize}
}\end{tcolorbox}

The other researcher would know that columns had been deleted, but insofar as the list only includes one column with a non-standard Qualtrics label, \textit{name}, she would have little reason for concern that these deletions had occurred for deceptive reasons. A far greater cause for concern would be the fact that at least one entry in the \textit{color} column had been changed.

\subsection{Computation and Storage Architecture}

The web-based application consists of two main components: a user interface and a backend server. The user interface facilitates user interactions, while the backend handles data processing, computation, and integration with external data sources and storage. The primary data storage operation occurs during the digest mode, at which point the system processes survey data files, computes column hashes, and stores this information along with related survey and user metadata in an external database for future verification. Notably, the portal does not retain the original .csv files, mitigating any confidentiality concerns. The external database can be hosted on cloud services like AWS S3, MS SQL Server, or other storage solutions as needed.

\begin{figure}
\caption{\label{fig:architecture}Data-NoMAD's Computation and Storage Architecture}
\begin{center}
    \includegraphics[width=0.8\linewidth]{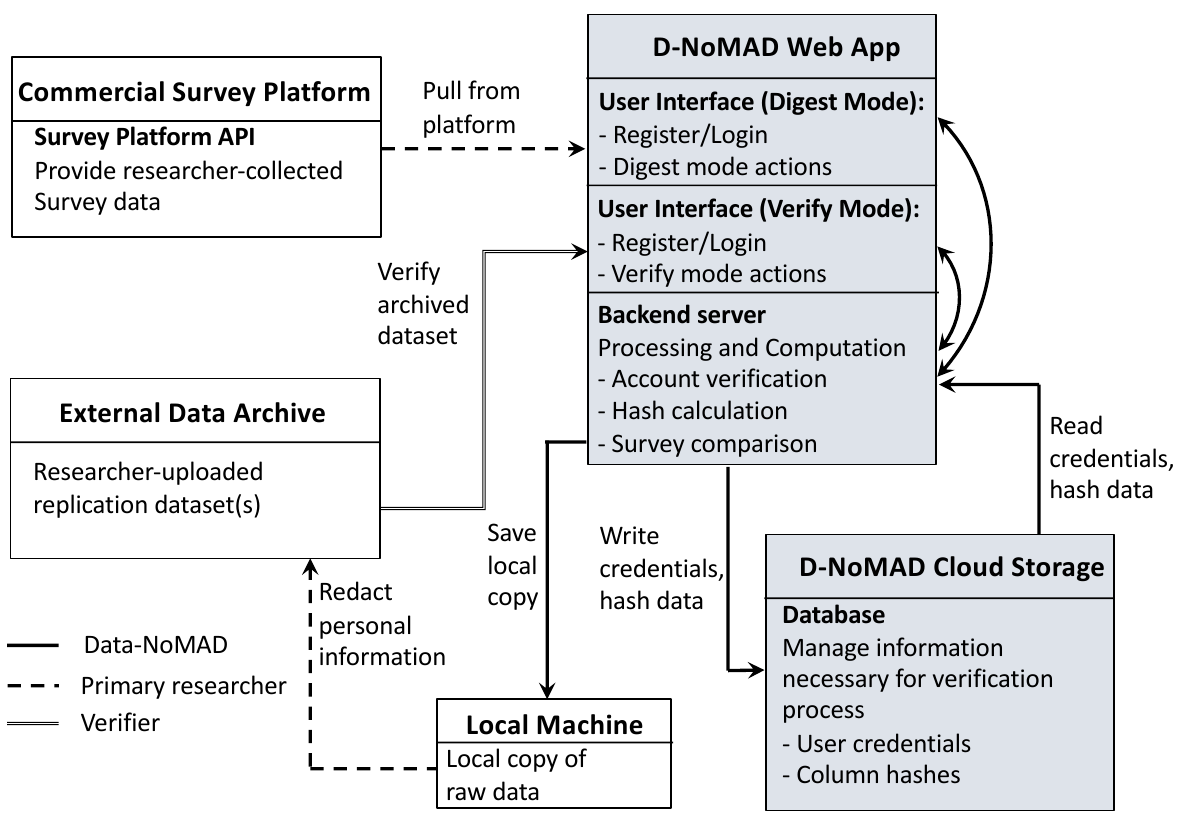}
\end{center}
\textit{Shaded boxes denote the internal architecture while unshaded boxes refer to external data platforms that are part of the workflow.}
\end{figure}

Figure \ref{fig:architecture} captures Data-NoMAD's essential architecture. The internal components are the user interface, backend server, and cloud storage. In addition to the researcher's own local computer,  Data-NoMAD also interfaces with two types of external services: survey platforms and external data archives.

\paragraph{User interface.} This is the front-facing part of the web app. The interface facilitates all user actions, including registration, login, downloading raw survey data, and uploading de-identified .csv files for comparison. It also acts as the interface through which users interact with the system.

\paragraph{Backend Server.} This serves as the core processing unit, responsible for key operations: 
\begin{itemize}
    \item  Authentication: Manages user registration, login, and email verification.
Database Connection: Ensures integration with the external database for efficient data operations
    \item  Data Processing: Handles tasks such as hashing IP addresses, comparing uploaded .csv files, and verifying data integrity
    \item  API Interaction: Connects with external commercial survey platforms to retrieve survey data based on specified identifiers
\end{itemize}

\paragraph{Cloud Storage} This encompasses all tasks related to storing, retrieving, and updating information, including:
\begin{itemize}
    \item Store User Information: Saves user credentials and verification details during registration.
    \item Store/Update Hashes: Computes and stores column hashes when processing survey data, ensuring that changes can be tracked and verified.
    \item Fetch Data: Retrieves stored user information, or existing column hashes for comparison during various processes.
\end{itemize}

The workflow involves two data platform components that are external to Data-NoMAD's architecture.
The first external component is the commercial survey platform. This is an external service that hosts the survey data collection. Data-NoMAD's backend server fetches data from these platforms via API calls guided by user-provided information (survey ID, API keys, etc.). The second component is the external data archive. This is an external service or website (often hosted by universities, journals, or individual researchers) on which anonymized data are stored for use by third parties.
third parties would typically access the archived data from such a service or website to then authenticate using Data-NoMAD's verify mode.

\subsection{Patching Vulnerabilities}

No method for ensuring data integrity is 100\% tamperproof. Data-NoMAD significantly mitigates the cell manipulation, column addition and deletion, and row addition problems that might be part of a strategy to create fraudulent data using filed downloaded by the analyst. Some vulnerabilities remain, and, importantly, are platform-specific. Here, we discuss these vulnerabilities and the functionality that already exists to address two of them, as well as what is required to fully address the third.

\paragraph{Importing bogus responses.} The Qualtrics and SurveyCTO platforms grant users with administrative privileges the ability to import responses from external data stored in .csv, .tsv or .txt format.\footnote{As of writing, the instructions for importing data were available at the following: for Qualtrics, \url{https://www.qualtrics.com/support/survey-platform/data-and-analysis-module/data/import-responses/}, and for SurveyCTO, 
\url{https://support.surveycto.com/hc/en-us/articles/360047406414-Working-between-servers-including-transferring-form-definitions-and-data-between-them}.} In Qualtrics, this practice leaves a tell-tale sign: for any rows that are imported, the corresponding entries in the ``Status'' field say ``Imported.'' Thus, anyone using Data-NoMAD in verify mode could see in an instant whether a dataset contained imported rows. Below, we discuss best practices for researchers seeking to appropriately combine datasets without arousing suspicion.

\paragraph{In-platform response editing.} Both Qualtrics and SurveyCTO have functionality that permit response editing by the researcher. Fortunately, it is possible on both to verifiably recover the \textit{unaltered} data -- in Qualtrics, through a parameter in the API call that discards any edits; and in SurveyCTO through the data selection mode.

\paragraph{Permanent row deletion.} In our view, the biggest remaining vulnerability is that Qualtrics and SurveyCTO platforms grant users with administrative privileges the ability to permanently delete responses without any log or retrievable pre-tampering cache file. 
\footnote{At the time of writing, instructions for such permanent deletion were available at the following: for Qualtrics, \url{https://www.qualtrics.com/support/survey-platform/data-and-analysis-module/data/recorded-responses/\#DeletingResponses},  and for SurveyCTO, \url{https://support.surveycto.com/hc/en-us/articles/360024883774-Purging-data-on-your-server}.} As far as we can discern, patching this vulnerability will require changing features of the survey platforms.
We hope to work with survey platforms to design a mode that disables the functionality that can give rise to these.

\subsection{Best practices}

Data-NoMAD can help researchers who want to demonstrate the authenticity of their research. 
This approach works best when coupled with other practices that help to ensure transparency and reproducibility:
\begin{enumerate}
    \item Pilot tests of survey modules, practice runs by investigators or research assistance, or other checks prior to deployment of the survey to actual respondents should ideally be done in a separate project file. That way, the  that is ultimately downloaded for analysis does not contain practice observations or other observations that need to be deleted. We recommend distinguishing the two projects with a simple naming convention such as ``project.test'' and ``project.fielded.'' The value of this approach is is to have a digested file that comes as close as possible to the file that is actually used for data analysis. 
    \item Before downloading the data through the digest mode, have the survey closed out so that it cannot receive new responses.  
    \item To the extent possible, disable functionalities on survey platforms that allow for the removal or manipulation of rows, columns, or cell values without record and prior to exporting data.  If the researcher is obligated to remove observations upon a subject's request (e.g., after a debrief), then this instruction could be incorporated as a ``delete requested'' variable that Data-NoMAD uses to automatically delete rows before hashing and to document the action as part of the Data-NoMAD output. 
    \item Use a flat plain text format (such as comma-, tab-, or pipe-separated values) for data exported from survey platforms and stored archivally. Statistical packages can all input this format, and as shown in the example above, such formats can store column metadata. 
    \item The \textit{only} file manipulation that should take place offline in creating an archival dataset is the deletion of identifying information. The exception is cases in which the research project uses a block of survey items that piggybacked on a larger survey and the researcher has valid reasons to exclude the broader universe of items from the project-specific research archive. In such cases, more extensive column deletions may be justified. Under those circumstances, researchers can bolster the credibility of their research by pre-registering the complete list of questions that will ultimately be used and archived.
    \item In general, we recommend against deleting nuisance columns until the ``data preparation'' stage. Because of its utility in safeguarding against bogus row imports, it is essential that the ``Status'' field in Qualtrics not be deleted.
    
    The ideal output from Data-NoMAD in verify mode applied to a stand-alone research project and associated archival dataset would look something like this:
\begin{tcolorbox}[width=\linewidth, sharp corners=all, colback=white!85!green]
{\sffamily
\singlespacing
\noindent Changes detected.
\begin{itemize}
    \item Removed columns: IPaddress, name 
\end{itemize}
}\end{tcolorbox}
    \item All other manipulations done to prepare the data for analysis -- row and column additions or deletions, data transformations, etc. -- should be done in a replicable ``data preparation'' script (e.g., a .R file in R or .do file in Stata) that is included in the replication archive, along with the survey ID necessary to use Data-NoMAD in verify mode. In the event that two or more datasets are combined to conduct the analysis, the separate files should be independently digested if possible and combined at this data preparation stage. All pertinent survey IDs should be provided with the replication materials to permit verification. 
    \item For convenience of third parties, the standard practice should be for the data preparation script to be separate from the analysis script. 
\end{enumerate}

\section{Conclusion}

This paper introduces an {\it ex-ante} strategy, called Data-NoMAD, for enhancing the credibility of the data employed in testing social science hypotheses.
At present, Data-NoMAD operates as a cloud-based app.
When in ``digest'' mode, researchers can use Data-NoMAD's SHA-256 encryption to create a reference ``digest'' file of their data. 
The researcher can then deposit their data, with identifying information removed, in a replication archive of their choosing. 
When in ``verify'' mode, a third party can obtain the data from the replication archive and check it against the digest stored by Data-NoMAD. 
Data-NoMAD can then indicate any row or column additions or column manipulations.  
The strategy significantly mitigates problems of manipulated responses and the deletion or addition of rows or columns that researchers could try to use to fabricate results.
We present toy examples in the text above, and an operational cloud-based version is currently available for demonstration use. 

The Data-NoMAD approach helps to avoid problems of ex-post auditing, including underprovision of auditing due to free-riding or the fact that would-be fraudsters can learn from past fraud disclosures. 
By reducing suspicion of fraud in the first place, Data-NoMAD reduces reliance on auditing and thus the likelihood of costly Type I errors that might emerge if an honest scholar is wrongfully accused of malfeasance. 

At present, Data-NoMAD is intended primarily for survey research rather than research involving observational or administrative data. In principle, however, many of the practices discussed above may be modified to apply to these applications. In a future paper we will discuss these modifications further. 

Data-NoMAD never has to store identifying information, it does not interfere with legitimate data manipulation practices, and it perfectly complements other research integrity practices, such as the posting of replication data and scripts and pre-registration.  
Data-NoMAD also encourages researchers to adopt other ``best practices'' for reproducible research. 
This includes making minimally tampered raw data available in replication archives with scripts that separately prepare and then analyze data. 
Vulnerabilities that remain include data editing or data importing that platforms permit before exporting data.   Further effort is needed in collaboration with such platform providers to close such loopholes and allow for researchers who want to demonstrate their integrity to do so in a convincing manner.

\bibliography{dNoMAD}

\end{document}